\documentclass[aps,prl,superscriptaddress,twocolumn,notitlepage]{revtex4-1}
\usepackage[latin1]{inputenc}
\usepackage{amsmath}
\usepackage{amsfonts}
\usepackage{amssymb}
\usepackage{graphicx}
\usepackage{epstopdf}
\graphicspath{ {images/} }
\usepackage{amsmath}
\usepackage[titletoc]{appendix}
\usepackage{mathrsfs}
\DeclareMathOperator{\sign}{sign}
\DeclareMathOperator{\erfi}{erfi}

\def \DD{\mathcal{D}}

\begin{document}

\title{Tunneling probe of fluctuating superconductivity in disordered thin films}

\author{David Dentelski}
\affiliation{Department of Physics, Bar-Ilan University, 52900, Ramat Gan Israel}
\affiliation{Center for Quantum Entanglement Science and Technology, Bar-Ilan University, 52900, Ramat Gan Israel}


\author{Aviad Frydman}
\affiliation{Department of Physics, Bar-Ilan University, 52900, Ramat Gan Israel}

\author{Efrat Shimshoni}
\affiliation{Department of Physics, Bar-Ilan University, 52900, Ramat Gan Israel}

\author{Emanuele G.~Dalla Torre}
\affiliation{Department of Physics, Bar-Ilan University, 52900, Ramat Gan Israel}
\affiliation{Center for Quantum Entanglement Science and Technology, Bar-Ilan University, 52900, Ramat Gan Israel}

\begin{abstract}
Disordered thin films close to the superconducting-insulating phase transition (SIT) hold the key to understanding quantum phase transition in strongly correlated materials. The SIT is governed by superconducting quantum fluctuations, which can be revealed for example by tunneling measurements. These experiments detect a spectral gap, accompanied by suppressed coherence peaks that do not fit the BCS prediction. To explain these observations, we consider the effect of finite-range superconducting fluctuations on the density of states, focusing on the insulating side of the SIT. We perform a controlled diagrammatic resummation and derive analytic expressions for the  tunneling differential conductance. We find that short-range superconducting fluctuations suppress the coherence peaks, even in the presence of long-range correlations. Our approach offers a quantitative description of existing measurements on disordered thin films and accounts for tunneling spectra with suppressed coherence peaks observed, for example, in the pseudo gap regime of high-temperature superconductors.
\end{abstract}

\maketitle
	    
\newpage
	
{\it Introduction --}
Superconducting thin films have attracted recently a lot of attention due to the possibility of observing a direct superconductor-to-insulator transition (SIT) \cite{lin2015superconductivity, thinfilms, thinfilms2}. The SIT is considered as an excellent example of a quantum phase transition \cite{sachdev2001quantum}: it occurs at temperature $T = 0$ and is driven by a non-thermal tuning parameter $g$.  Experimentally, the SIT can be driven by a wide variety of $g$, such as thickness \cite{strongin1970destruction, dynes1986breakdown, haviland1989onset, valles1992electron, merchant2001crossover, frydman2002universal, hadacek2004double, stewart2007superconducting, sacepe2008disorder, hollen2011cooper, baturina2011nanopattern,  poran2017quantum}, magnetic \cite{hadacek2004double,stewart2007superconducting, paalanen1992low, yazdani1995superconducting,gantmakher2000observation,sambandamurthy2004superconductivity,sambandamurthy2005experimental,steiner2005possible,baturina2005quantum,baturina2007quantum,crane2007fluctuations,vinokur2008superinsulator} or electric fields \cite{parendo2005electrostatic}, chemical composition \cite{mondal2011phase, higgs}, and disorder \cite{crane2007fluctuations, poran2011disorder}. Near the quantum critical point $ g = g_{c}$, the system is governed by quantum fluctuations \cite{shimshoni1998transport, bouadim2011single, dubi2007nature, erez2010thermal, erez2013proposed}, and cannot be described in terms of classical Ginzburg-Landau theories \cite{cohen1969effect,abrahams1970effect,di1990superconductive,varlamov1999role}.

In the vicinity of the SIT, experiments show an intriguing behavior of the superconducting energy gap $\Delta$. Traditionally, $\Delta$ is determined by fitting the tunneling conductivity to a phenomenological extension of the BCS theory, which takes into account an effective energy broadening $\Gamma$ and is known as the Dynes formula \cite{Dynes},
		\begin{align}\label{Dynes}
		\dfrac{dI}{dV}(V) = {\rm Re}\left[ \dfrac{V-i\Gamma}{\sqrt{(V-i\Gamma)^2-\Delta^2}}\; \right].
		\end{align}	
This procedure is very useful in extracting values of $\Delta$ as a function of temperature for relatively clean superconductor. In these materials, $\Delta$ shrinks to zero as $T$ approaches the critical temperature $T_c$, in agreement to the predictions of the BCS theory \cite{BCS}. In contrast, in disordered thin films, tunneling experiments have revealed that $\Delta$ smoothly evolves across the transition \cite{barber1994tunneling,sherman2012measurement}, and through $T_c$ \cite{sacepe2011localization}.

In addition to this non-conventional behavior of $\Delta$, superconducting thin films show a significant deviation of the experimentally measured DOS from Eq.~(\ref{Dynes}), due to a considerable suppression of the coherence peaks at the gap edges \cite{sacepe2011localization, sherman2012measurement}. A similar disagreement was observed in high-temperature superconductors \cite{zasadzinski1992tunneling, dynes1994order}. This discrepancy can be accounted for by assuming the two $\Gamma$'s that appear in the numerator and in the denominator of Eq.~(\ref{Dynes}) to be different \cite{norman1998phenomenology, norman2001temperature,norman2007modeling}, but this approach lacks physical insight. 

In this Letter we suggest an alternative approach which relates the experimental findings to a well defined theoretical model.  
Instead of considering an effective energy broadening, we include superconducting fluctuations by postulating a bosonic field  $\Delta(\textbf{r},t)$ with a finite correlation length. By summing the contributions of short range and long range fluctuations, we obtain an excellent agreement with experimental curves on the insulating side of the SIT. Our results demonstrate that there are two important length scales: one is the effective size of a superconducting island, $\xi_{\rm sc}$, and the other is the typical size of quantum fluctuations $\xi_{\rm fluc}$, which diverges at the SIT.


{\it Superconducting fluctuations --}
We begin our analysis by introducing the Hamiltonian $H=H_{0}+H_{\Delta}$, where $H_{0} = \sum_{\textbf{k}, \sigma} \varepsilon_{\textbf{k}} c^{\dagger}_{\textbf{k}, \sigma}c_{\textbf{k}, \sigma}$ describes free electrons (quasiparticles) with a Fermi surface at $\varepsilon_{\textbf{k}}=0$. Following Ref.~ \cite{abrahams1970effect,di1990superconductive}, we represent superconducting fluctuations by a randomly fluctuating bosonic field $\Delta(\textbf{r}, t)$ coupled to the fermions by
\begin{align}
	H_{\Delta} = \Delta(\textbf{r},t)c_{\uparrow}^\dagger (\textbf{r},t) c_{\downarrow}^\dagger(\textbf{r},t) + {\rm H. c.}\;.
	\label{eq:H}
	\end{align}	
In insulating samples, the superconducting order parameter averages to zero, $\langle\Delta(\textbf{r},t)\rangle = 0$. We model finite-range superconducting fluctuations by a free field with two-point correlations 
\begin{align}
C(\textbf{r}-\textbf{r'},t-t') = \langle \Delta(\textbf{r},t) \Delta^{*}(\textbf{r'},t')\rangle\;.
\end{align}
The function $C$ describes the decay of the superconducting correlations due to the interplay between disorder and interactions, and tends to zero at long distances and large times.

We now derive the relation between  $C(\textbf{r}-\textbf{r}',t-t')$ and tunneling measurements. The tunneling differential conductivity is proportional to \cite{mahan2000many}
\begin{align}\label{IV}
\dfrac{dI}{dV}(V) \propto \int_{-\infty}^{\infty} d\omega ~\rho(V+\omega) f'(\omega)\;.
\end{align}
Here $V$ is the voltage bias, $f'(\omega)=df/d\omega$ is the derivative of the Fermi-Dirac distribution function, and $\rho(\omega)$ is the density of states (DOS) of the sample. Eq.(\ref{IV}) assumes a constant DOS of the tip, and at $T=0$, simply reduces to ${dI}/{dV}\propto \rho(V)$. Within the Green's function formalism in Nambu space \cite{altland2006condensed}, $\rho(\omega) =  -({1}/{\pi})\langle {\rm Im} \left\lbrace{\rm Tr}\left[ \sum_{\textbf{k}} G^{ret}(\textbf{k},\omega)\right]\right\rbrace\rangle$, where $G^{ret} (\textbf{k},\omega)$ is the retarded Green's function, ${\rm Tr}$ is the trace in Nambu space, $\rm Im$ is the imaginary part, and $\langle...\rangle$ implies an average over the fluctuations of the superconducting field $\Delta (\textbf{r}, t)$. 

\begin{figure} [t] 
	\includegraphics[scale=0.5]{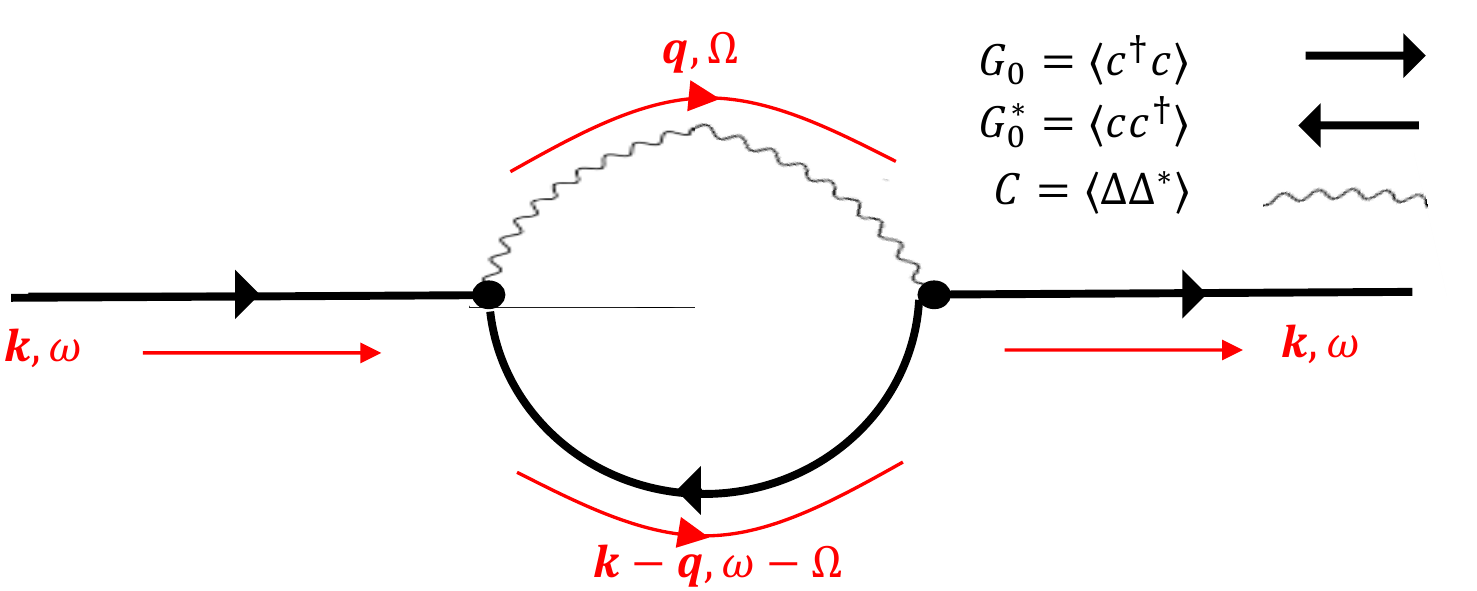}
	\centering
	\caption{One loop diagram - second order perturbation in $\Delta(\textbf{r},t)$. The black thick arrows represents charge (i.e. right arrow for a particle and left arrow for a hole),  while the thin red ones represents momentum and energy.}
	\label{Fyenman}
\end{figure}

Our first step involves a Dyson resummation of the one-loop contributions shown in Fig.~\ref{Fyenman}, whose two vertices represent the coupling term (\ref{eq:H}). By performing a trace over particle and hole contributions, we find
\begin{align}\label{Green function}
{\rm Tr}\left[ G^{ret}(\textbf{\textbf{k}},\omega)\right] = \dfrac{\omega+i0^{+}}{(\omega+i0^{+})^2-\varepsilon_{\textbf{k}}^2-\mathcal{D}^2(\textbf{k},\omega)}\;.
\end{align} 
Here we defined the pairing-fluctuations function $\DD$ as
\begin{align}\label{pairing1}
\mathcal{D}^2(\textbf{k},\omega) = \int d^2\textbf{q} \int d\Omega \dfrac{\omega+i0^{+}+\varepsilon_{\textbf{k}}}{i\Omega +\omega +i0^{+}+\varepsilon_{\textbf{k}-\textbf{q}}}C(\textbf{q}, \Omega),
\end{align}
with $C(\textbf{q}, \Omega) = \int d^2 \textbf{r}~ dt~ C(\textbf{r}, t)~e^{i\textbf{q} \cdot \textbf{r}-i\Omega t}$. 

Because the Green's function in Eq.~(\ref{Green function}) is strongly peaked at ${\bf k}={\bf k}_F$, we can approximate the density of states as
\begin{align}\label{DOS}
\rho(\omega) =  {\rm Re} \left[ \frac {\omega}{\sqrt{\omega^2 -\mathcal{D}^2(\omega)}}\right]\;,
\end{align}
where we defined $\DD(\omega)\equiv\DD({\bf{ k_F}},\omega)$. 

Eq.~(\ref{DOS}) is analogous to the Dynes equation (\ref{Dynes}), but involves the frequency dependent $\DD(\omega)$ instead of the quasiparticle lifetime $\Gamma$. Note that if the correlation function $C(\textbf{r}, t)$ does not decay in space and time (i.e. the BCS limit), its Fourier transform is $C(\textbf{q},\Omega)\sim\delta(\textbf{q})\delta(\Omega)$. In this case, Eq.~(\ref{pairing1}) yields a frequency independent $\DD^2(\omega)=\Delta_0^2$, and one recovers the well-known result. 

Our approach has some similarities to Refs.~\cite{cohen1969effect,abrahams1970effect,di1990superconductive}, where superconducting fluctuations with a finite lifetime were considered as well. These authors were interested in the thermal regime $T\gg T_c$, where superconducting fluctuations are weak and lead to small deviations of the density of states. As a consequence, their approximation scheme does not recover the diverging density of states predicted by BCS. The Dyson resummation employed in the present work allows us to consider strong superconducting fluctuations and get closer to the SIT. See also Ref.~\cite{burmistrov2016local} for a microscopic model describing the effect of superconducting fluctuations close to the SIT and their effects on the DOS.



\begin{figure*} [t]
\centering
\begin{tabular}{p{4cm} p{4cm} p{4cm} p{4cm}}
   \ \ \ \ \  \begin{tabular}{c} (a) Long range (LR) \\ $v_{F}q_{0}/\Delta_{0} = 0.1$ \end{tabular} &    \ \ \ \ \  \ \ \begin{tabular}{c}   (b)  Finite Range \\ $v_{F}q_{0}/\Delta_{0} = 1$ \end{tabular}&  \ \ \ \ \begin{tabular}{c}
  (c) Short range (SR) \\ $v_{F}q_{0}/\Delta_{0} = 10$ \end{tabular}&        \begin{tabular}{c}(d) Quasi-long range (QLR) \\ $v_{F}q_{0}/\Delta_{0} = 0.03$\end{tabular}
\end{tabular}
\vspace{-0.5cm}
\hspace{-1cm}
\includegraphics[scale=0.8,trim=0 0 0 20,clip]{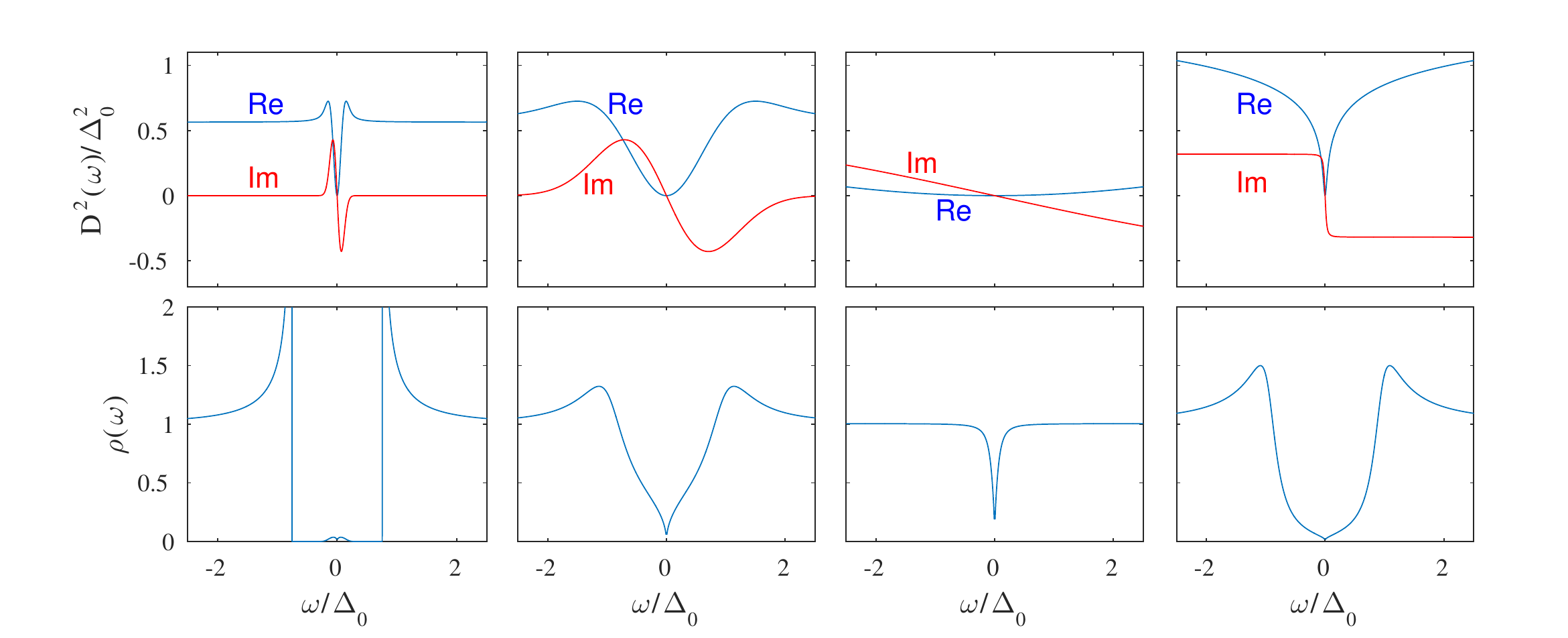}

\caption{Upper panel: (a-c) Real and imaginary part of Eq.~(\ref{longD}), for different values of $v_F q_0/\Delta_0$. (d) Real and imaginary part of Eq.~(\ref{qlrD}). Lower panel: the corresponding density of states, Eq.~(7).}
	
	\centering
	\label{SigmaShort}
\end{figure*}

In what follows, for simplicity, we will generically assume that the correlations are time independent, i.e.
\begin{align}
 C(\textbf{q},\Omega) = C(\textbf{q})\delta(\Omega).
\end{align}
This assumption is justified if the collective-mode velocity $v$ is much smaller than the Fermi velocity $v_{F}$: See the supplemental information for more details about the validity of this approximation. Assuming a quadratic dispersion relation $\varepsilon_\textbf{k} = (k^2-k_{F}^2)v_F/2k_F$  and assuming that $C(\textbf{q})$ decays to zero at $q\sim k_{F}$, we obtain
\begin{align}\label{SigmaS}
\mathcal{D}^2(\omega) = \int d^2\textbf{q}\dfrac{1}{\omega -v_{F}q_{x}}C(\textbf{q}).
\end{align}

Eqs.~(\ref{DOS}) and (\ref{SigmaS}) are the key theoretical results of our analysis, and we will now use them to model the density of states of disordered superconductors under various assumptions on the form of $C(\textbf{q})$. In the following, the films are assumed to be thin enough such that both $\textbf{k}$ and $\textbf{q}$ can be treated as two dimensional.

{\it Short range vs. long range --}
In order to understand the effects of superconducting fluctuations on the density of states, we consider correlation functions decaying over a typical inverse length scale $q_{0}$. This quantity can be associated with the average size of superconducting islands in the granular materials, and with an emergent electronic granularity of amorphous materials \cite{shimshoni1998transport, bouadim2011single, dubi2007nature, erez2010thermal, erez2013proposed}. In the specific case of  $C(\textbf{q}) \sim \exp \left(-{q^2}/{q_{0}^2}\right)$,  we can analytically solve the integral in Eq.~(\ref{SigmaS}) to find
\begin{align}\label{longD}
\mathcal{D}^2(\omega) =\Delta_0^2 \frac{1}{v_{F}q_{0}} \exp\left(-\dfrac{\omega^2}{v_{F}^2q_{0}^2}\right) \omega \left[\erfi\left(\dfrac{\omega}{v_{F}q_{0}}\right)-i\right]\;.
\end{align}
Here $\erfi$ is the imaginary error function, which is a real function. The real and imaginary parts of Eq.~(\ref{longD}) are shown in the upper panel of Fig.~\ref{SigmaShort}(a-c) for different values of $v_{F}q_{0}/\Delta_{0}$, and the corresponding DOS in the lower panel. Note that the real part of  $\mathcal{D}^2(\omega)$ closely resembles the local density of states $\rho(\omega)$, but these two quantities have a different physical meaning: the former actually needs to be substituted in Eq.~(\ref{DOS}) to deliver the latter. We observe that both $\rm Re[\mathcal{D}^2]$ and $\rm Im[\mathcal{D}^2]$ are peaked at a typical energy scale $v_{F} q_{0}$.

The effect of superconducting fluctuations on the DOS changes dramatically, depending on the ratio between $v_{F} q_0$ and $\Delta_{0}$. Let us consider two extreme cases, which we denote by long range (LR) and short range (SR), respectively. The former case occurs for $v_F q_{0} \sim v_F \xi_{\rm fluc}^{-1} \ll \Delta_{0}$. In this case $\DD^2(\omega)$ is approximately constant and we recover the BCS limit (see Fig.~\ref{SigmaShort}(a)). On the other hand, when $v_F q_{0} \sim v_F\xi_{\rm sc}^{-1}  \gg \Delta_{0}$, the fluctuations are short ranged. In this regime
\vspace{-0.2cm}
\begin{align}
\mathcal{D}^2 (\omega)\approx - i \omega \gamma\label{eq:DDSR}\vspace{-0.5cm}
\end{align}
is purely imaginary, and the DOS shows a deep without coherence peaks (see Fig.~\ref{SigmaShort}(c)). The distinction between LR and SR fluctuations does not depend on the specific choice of $C(\textbf{q})$, and can be related to the Anderson limit of superconductivity \cite{Anderson, seibold2015amplitude}. The crossover between these two regimes occurs when $\Delta_0$ is of the order of the typical energy level spacing of a  superconducting island of size $1/q_0$ (the superconducting correlation length), i.e. $\Delta_0\sim v_F q_0$.

In the vicinity of the SIT, superconducting fluctuations are described by a universal critical theory, which in its simplest form, is given by $C(\textbf{q}) \sim 1/(q^2 +q_{0}^2)$ (See Ref.~\cite{sondhi1997continuous}), where $q_{0} = 1/\xi_{\rm fluc}$ tends to zero at the transition. In this case, which we denote as quasi long range (QLR), we can again solve analytically Eq.~(\ref{SigmaS}) to obtain
\begin{align}\label{qlrD}
\mathcal{D}^2(\omega) = \frac{\Delta_{0}^{2}}{\pi^2}\dfrac{\omega}{\sqrt{v_{F}^2q_{0}^2+\omega^2}} \left[ \ln\left(\dfrac{\sqrt{v_{F}^2q_{0}^2+\omega^2}+\omega}{\sqrt{v_{F}^2q_{0}^2+\omega^2}-\omega}\right)-i\pi\right]\;.
\end{align}
As shown in Fig.~\ref{SigmaShort}(d), for $q_0\to0$, the real part of Eq.~(\ref{qlrD}) diverges logarithmically, while the imaginary part is proportional to ${\rm sign}(\omega)$. The resulting DOS resembles the LR situation and is very weakly dependent on the infrared cutoff $q_{0}$. As we will see below, QLR superconducting fluctuations give a better description of the experiment than true LR correlations.

\begin{figure}[t]
	\vspace{-0.7cm}
	\includegraphics[]{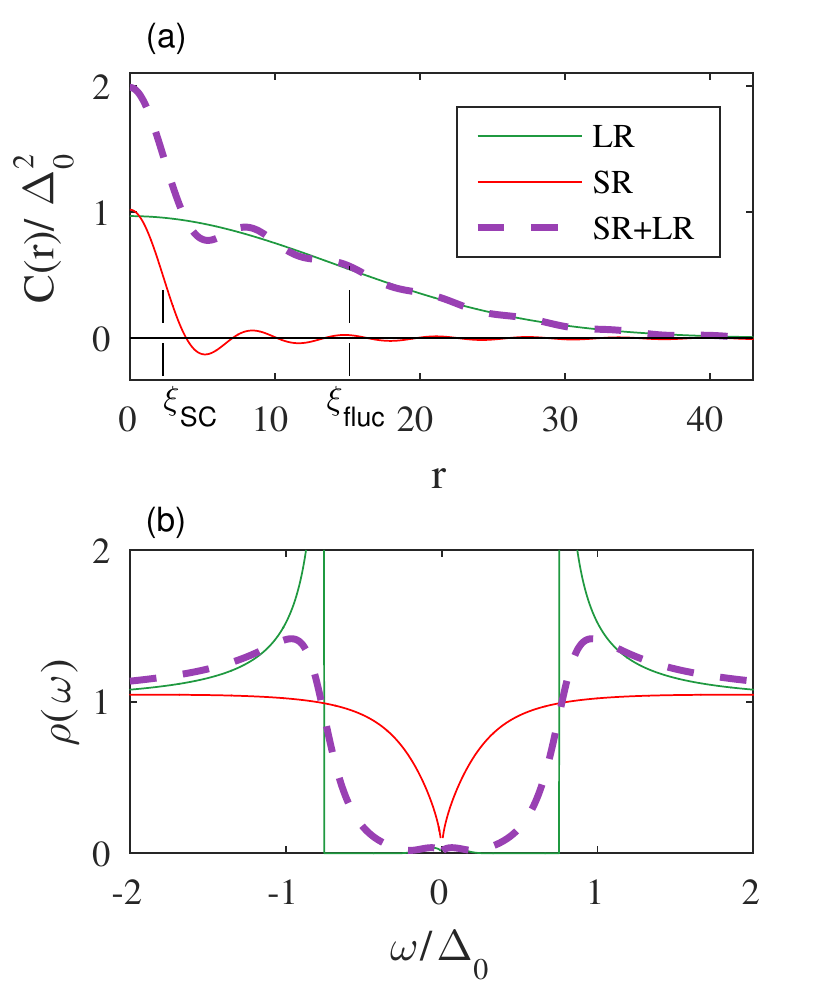}
	\vspace{-0.5cm}
	\caption{(a) Normalized spatial correlations of the superconducting fluctuations, $C(r)$; (b) Corresponded density of states $\rho(\omega)$. Note that adding short range fluctuations (SR, $v_F q_0/\Delta_{0}=3$) suppresses the peaks, even in the presence of long range superconducting correlations (LR, $v_F q_0/\Delta_0=0.1$).} \label{ShortRangeEffect}

\end{figure}

{\it Comparison with experiments --} In actual materials, one generically expects to find some combination of superconducting fluctuations with both long range and short range correlations. The former are universal and determine the emergent properties of the material, while the latter depend on the details of the materials and are often neglected. In contrast to this common practice, we find that short range correlation strongly affect the density of states (see  Fig.~\ref{ShortRangeEffect}): although the correlation functions denoted by LR and SR+LR have the same asymptotic behavior (subplot (a)), the corresponding DOS are very different (subplot (b)).

\begin{figure} [t]
	\includegraphics[scale=0.82]{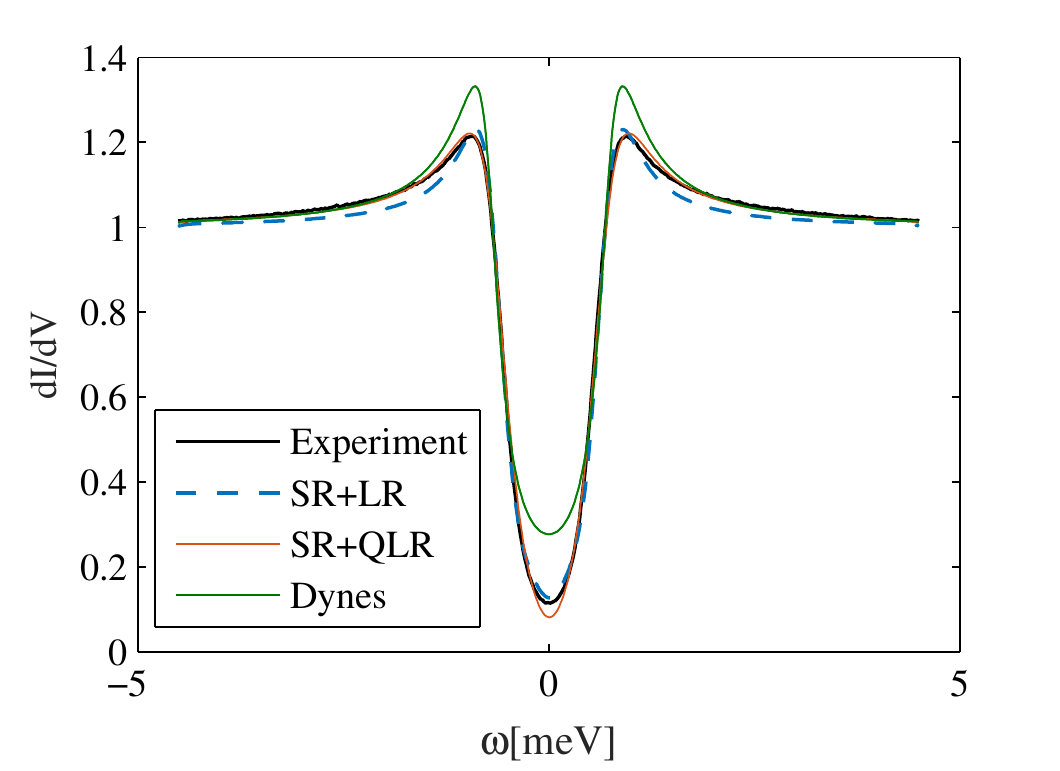}
	\centering
	\caption{Comparison between theoretical predictions and actual measurements performed on an insulating thin film close to the SIT. The best theoretical fits for each curve is presented at table \ref{table}). The best fit to the experiment is given by SR+QLR.}
	\label{DOSforLRandSR}
\end{figure}

We now compare our theoretical calculations with the tunneling measurement of Ref.~\cite{sherman2012measurement}, performed on an InO film at $T=1K$ at the insulating side of the SIT. Note that in order to isolate the superconducting contribution to the DOS, the experimental raw data was normalized by the tunneling spectra at high magnetic field. The results of our analysis are summarized in Table \ref{table} and Fig.~\ref{DOSforLRandSR}, where we show the best fitting parameters and the minimal normalized $\chi^{2}$-distribution between theory and experiment. We generically find that the sum of short range and long range superconducting fluctuations is required to obtain a good fit. Furthermore, a detailed analysis reveals that the long range part is best described by Eq.~(\ref{qlrD}) (QLR, $\chi^2=0.011$), rather than Eq.~(\ref{longD}) (LR, $\chi^2=0.022$), in agreement with the expected critical behavior of the SIT \cite{sondhi1997continuous}.


{\it Discussion --} In this Letter we studied the effects of superconducting fluctuations on the tunneling conductivity of disordered thin films, focusing on the insulating side of the SIT. The common approach, known as Dynes formula, relies on a phenomenological parameter $\Gamma$ that describes the inverse lifetime of the quasiparticles. In this study, we showed that the experiments are better fit by a theory of free electrons, coupled to superconducting fluctuations with finite-range correlations. By using a controlled diagrammatic approach, we derived a simple expression that connects the correlations of the superconducting fluctuations to the tunneling spectra, Eqs.~(\ref{pairing1}) and (\ref{DOS}).  This result has potential applications that go beyond the present study, including quantum as well as classical superconducting phase transitions. Our analytical results show that, generically, short-range fluctuations lead to tunneling spectra with reduced or absent coherence peaks, even in the presence of long-range superconducting correlations.
\begin{table}[t]
	\begin{tabular}{|c|  c| c | c| } 
		\hline
		& DOS & best fit [meV] & $\chi^2$ \\
		\hline	
		Dynes&  Eq.(\ref{Dynes}) &  \begin{tabular}{c} $\Delta_0=0.73 $\vspace{-.1cm} \\ $\Gamma=0.16$ \end{tabular} & 0.046  \\
		\hline
		SR+LR & \begin{tabular}{c} Eq. (\ref{DOS}) with \\ $\DD^2 (\omega)=$ Eqs.~(\ref{eq:DDSR})+(\ref{longD}) \end{tabular} & \begin{tabular}{c}$\Delta_0=0.67$\vspace{-.1cm} \\ $\gamma=0.22$\vspace{-.1cm} \\ $v_{F}q_{0} = 0.22$\end{tabular} & 0.022 \\
		\hline
		SR+QLR & \begin{tabular}{c} Eq. (\ref{DOS}) with \\ $\DD^2 (\omega)$ = Eqs.~(\ref{eq:DDSR})+(\ref{qlrD})  \end{tabular} & \begin{tabular}{c}$\Delta_0 =0.69 $\vspace{-.1cm} \\ $\gamma= 0.11$\vspace{-.1cm} \\ $v_{F}q_{0} = 0.022  $\end{tabular} &  0.011 \\
		\hline
	\end{tabular} 
	\caption{Comparison between different correlation function, i.e. Dynes, SR+LR and SR+QLR.} 
	\label{table}
	\vspace{-0.5cm}
\end{table}

By comparing our analytic expressions to experimental measurements, we find that in disordered thin films, the superconducting fluctuations are given by the sum of two components. The long range component is associated with universal fluctuations close to the SIT quantum critical point, characterized by a diverging length scale $\xi_{\rm fluc}$. Accordingly, the experimental data is best fit by a critical theory with $q_0\sim 1/\xi_{\rm fluc} \ll k_F$ (see the last row of Table \ref{table}, taking into account that $v_F k_F\sim 1eV)$. In contrast, the short range component is determined by the microscopic details of the material. Specifically, short range correlations are expected to play a predominant role in amorphous materials, where Cooper pairs are localized by disorder. In granular materials, on the other hand, the superconducting correlations decay over a much longer range, set by the typical scale of the grains. This distinction can explain why the Dynes formula (\ref{Dynes}) fits well experiments on granular Pb films \cite{merchant2001crossover}, but does not fit amorphous InO films \cite{sherman2012measurement}. The distinction between short range and long range fluctuations can bridge the long standing controversy between the fermionic and bosonic approach to the SIT \cite{seibold2015amplitude}.
On a broader prospective, our approach contributes to the understanding of puzzling spectrometric experiments in unconventional superconductors. Specifically, we find that although SR fluctuations contribute to the local superconducting gap, they generically lead to a tunneling spectrum with suppressed coherence peaks. When only SR fluctuations are present, the spectrum becomes v-shaped (see Fig. \ref{SigmaShort}(c)), in analogy to the experimental observations in the pseudo-gap regime of underdoped cuprates (see for example Refs. \cite{norman1998phenomenology, norman2001temperature, boyer2007imaging,fujita2011spectroscopic}).


{\it Acknowledgments --}
We thank Assa Auerbach, Igor Burmistrov, Tatiana Baturina, Oriya Eizenberg and Eugene Demler for useful discussions. This work is supported by the Israel Science Foundation, grant No. 1452/14 (E.G.D.T. and D.D.) and No. 231/14 (E.S.). A.F. acknowledges the support of the GIF foundation grant
I-1250-303.10/2014. D.D. thanks the Klein family for a fellowship in memory of Prof. Michael Klein.


\begin{thebibliography}{10}
	\expandafter\ifx\csname url\endcsname\relax
	\def\url#1{\texttt{#1}}\fi
	\expandafter\ifx\csname urlprefix\endcsname\relax\def\urlprefix{URL }\fi
	\providecommand{\bibinfo}[2]{#2}
	\providecommand{\eprint}[2][]{\url{#2}}
	
	\bibitem{lin2015superconductivity}
	\bibinfo{author}{Lin, Y.-H.}, \bibinfo{author}{Nelson, J.} \&
	\bibinfo{author}{Goldman, A.}
	\newblock \bibinfo{title}{Superconductivity of very thin films: The
		superconductor--insulator transition}.
	\newblock \emph{\bibinfo{journal}{Physica C: Superconductivity and its
			Applications}} \textbf{\bibinfo{volume}{514}}, \bibinfo{pages}{130}
	(\bibinfo{year}{2015}).
	
	\bibitem{thinfilms}
	\bibinfo{author}{{Goldman}, A.~M.} \& \bibinfo{author}{{Markovic}, N.}
	\newblock \bibinfo{title}{{Superconductor-insulator transitions in the
			two-dimensional limit}}.
	\newblock \emph{\bibinfo{journal}{Physics Today}}
	\textbf{\bibinfo{volume}{51}}, \bibinfo{pages}{39} (\bibinfo{year}{1998}).
	
	\bibitem{thinfilms2}
	\bibinfo{author}{Markovi\ifmmode~\acute{c}\else \'{c}\fi{}, N.},
	\bibinfo{author}{Christiansen, C.} \& \bibinfo{author}{Goldman, A.~M.}
	\newblock \bibinfo{title}{Thickness-magnetic field phase diagram at the
		superconductor-insulator transition in 2d}.
	\newblock \emph{\bibinfo{journal}{Phys. Rev. Lett.}}
	\textbf{\bibinfo{volume}{81}}, \bibinfo{pages}{5217} (\bibinfo{year}{1998}).
	
	\bibitem{sachdev2001quantum}
	\bibinfo{author}{Sachdev, S.}
	\newblock \emph{\bibinfo{title}{Quantum Phase Transitions}}
	(\bibinfo{publisher}{Cambridge University Press}, \bibinfo{year}{2001}).
	
	\bibitem{strongin1970destruction}
	\bibinfo{author}{Strongin, M.}, \bibinfo{author}{Thompson, R.},
	\bibinfo{author}{Kammerer, O.} \& \bibinfo{author}{Crow, J.}
	\newblock \bibinfo{title}{Destruction of superconductivity in disordered
		near-monolayer films}.
	\newblock \emph{\bibinfo{journal}{Physical Review B}}
	\textbf{\bibinfo{volume}{1}}, \bibinfo{pages}{1078} (\bibinfo{year}{1970}).
	
	\bibitem{dynes1986breakdown}
	\bibinfo{author}{Dynes, R.~C.}, \bibinfo{author}{White, A.~E.},
	\bibinfo{author}{Graybeal, J.~M.} \& \bibinfo{author}{Garno, J.~p.}
	\newblock \bibinfo{title}{Breakdown of eliashberg theory for two-dimensional
		superconductivity in the presence of disorder}.
	\newblock \emph{\bibinfo{journal}{Physical Review Letters}}
	\textbf{\bibinfo{volume}{57}}, \bibinfo{pages}{2195} (\bibinfo{year}{1986}).
	
	\bibitem{haviland1989onset}
	\bibinfo{author}{Haviland, D.~B.}, \bibinfo{author}{Liu, Y.} \&
	\bibinfo{author}{Goldman, A.~M.}
	\newblock \bibinfo{title}{Onset of superconductivity in the two-dimensional
		limit}.
	\newblock \emph{\bibinfo{journal}{Physical Review Letters}}
	\textbf{\bibinfo{volume}{62}}, \bibinfo{pages}{2180} (\bibinfo{year}{1989}).
	
	\bibitem{valles1992electron}
	\bibinfo{author}{Valles~Jr, J.~M.}, \bibinfo{author}{Dynes, R.~C.} \&
	\bibinfo{author}{Garno, J.~P.}
	\newblock \bibinfo{title}{Electron tunneling determination of the
		order-parameter amplitude at the superconductor-insulator transition in 2d}.
	\newblock \emph{\bibinfo{journal}{Physical Review Letters}}
	\textbf{\bibinfo{volume}{69}}, \bibinfo{pages}{3567} (\bibinfo{year}{1992}).
	
	\bibitem{merchant2001crossover}
	\bibinfo{author}{Merchant, L.}, \bibinfo{author}{Ostrick, J.},
	\bibinfo{author}{Barber~Jr, R.~P.} \& \bibinfo{author}{Dynes R.~C}
	\newblock \bibinfo{title}{Crossover from phase fluctuation to amplitude-dominated superconductivity: A model system}.
	\newblock \emph{\bibinfo{journal}{Physical Review B}}
	\textbf{\bibinfo{volume}{63}}, \bibinfo{pages}{134} (\bibinfo{year}{2001}).
	
	
	\bibitem{frydman2002universal}
	\bibinfo{author}{Frydman, A.}, \bibinfo{author}{Naaman, O.} \&
	\bibinfo{author}{Dynes, R.~C.}
	\newblock \bibinfo{title}{Universal transport in two-dimensional granular
		superconductors}.
	\newblock \emph{\bibinfo{journal}{Physical Review B}}
	\textbf{\bibinfo{volume}{66}}, \bibinfo{pages}{052509}
	(\bibinfo{year}{2002}).
	
	\bibitem{hadacek2004double}
	\bibinfo{author}{Hadacek, N.}, \bibinfo{author}{Sanquer, M.} \&
	\bibinfo{author}{Vill{\'e}gier, J.~C.}
	\newblock \bibinfo{title}{Double reentrant superconductor-insulator transition
		in thin tin films}.
	\newblock \emph{\bibinfo{journal}{Physical Review B}}
	\textbf{\bibinfo{volume}{69}}, \bibinfo{pages}{024505}
	(\bibinfo{year}{2004}).
	
	\bibitem{stewart2007superconducting}
	\bibinfo{author}{Stewart, M.~D.}, \bibinfo{author}{Yin, A.}, \bibinfo{author}{Xu,
		J.} \& \bibinfo{author}{Valles, J.~M.}
	\newblock \bibinfo{title}{Superconducting pair correlations in an amorphous
		insulating nanohoneycomb film}.
	\newblock \emph{\bibinfo{journal}{Science}} \textbf{\bibinfo{volume}{318}},
	\bibinfo{pages}{1273} (\bibinfo{year}{2007}).
	
	\bibitem{sacepe2008disorder}
	\bibinfo{author}{Sac{\'e}p{\'e}, B.}, \bibinfo{author}{Chapelier, C.}, \bibinfo{author}{Baturina, T.~I.}, \bibinfo{author}{Vinokur, V.~M.}, \bibinfo{author}{Baklanov, M.~R.} \& \bibinfo{author}{Sanquer, M.}
	\newblock \bibinfo{title}{Disorder-induced inhomogeneities of the
		superconducting state close to the superconductor-insulator transition}.
	\newblock \emph{\bibinfo{journal}{Physical Review Letters}}
	\textbf{\bibinfo{volume}{101}}, \bibinfo{pages}{157006}
	(\bibinfo{year}{2008}).
	
	\bibitem{hollen2011cooper}
	\bibinfo{author}{Hollen, S.~M.}, \bibinfo{author}{Nguyen, H.~Q}, \bibinfo{author}{Rudisaile, E.}, \bibinfo{author}{Stewart, M.~D.}, \bibinfo{author}{ Shainline,J.}, \bibinfo{author}{Xu, J.~M} \& \bibinfo{author}{Valles, J.~M.}
	\newblock \bibinfo{title}{Cooper-pair insulator phase in superconducting
		amorphous bi films induced by nanometer-scale thickness variations}.
	\newblock \emph{\bibinfo{journal}{Physical Review B}}
	\textbf{\bibinfo{volume}{84}}, \bibinfo{pages}{064528}
	(\bibinfo{year}{2011}).
	
	\bibitem{baturina2011nanopattern}
	\bibinfo{author}{Baturina, T.~I.}, \bibinfo{author}{Vinokur, V.~M.}, \bibinfo{author}{Mironov, A.~Yu.}, \bibinfo{author}{Chtchelkatchev, N.~M.}, \bibinfo{author}{Nasimov, D.~A.} \& \bibinfo{author}{Latyshev, A.~V.}
	\newblock \bibinfo{title}{Nanopattern-stimulated superconductor-insulator
		transition in thin tin films}.
	\newblock \emph{\bibinfo{journal}{EPL (Europhysics Letters)}}
	\textbf{\bibinfo{volume}{93}}, \bibinfo{pages}{47002} (\bibinfo{year}{2011}).
	
	\bibitem{poran2017quantum}
	\bibinfo{author}{Poran, S.}, \bibinfo{author}{Nguyen-Duc, T.},\bibinfo{author}{Auerbach, A.}, \bibinfo{author}{Dupuis, N.}, \bibinfo{author}{Frydman, A.} \& \bibinfo{author}{Bourgeois, O.}
	\newblock \bibinfo{title}{Quantum criticality at the superconductor-insulator
		transition revealed by specific heat measurements}.
	\newblock \emph{\bibinfo{journal}{Nature Communications}}
	\textbf{\bibinfo{volume}{8}}, \bibinfo{pages}{14464} (\bibinfo{year}{2017}).
	
	\bibitem{paalanen1992low}
	\bibinfo{author}{Paalanen, M.~A.}, \bibinfo{author}{Hebard, A.~F.} \&
	\bibinfo{author}{Ruel, R.~R.}
	\newblock \bibinfo{title}{Low-temperature insulating phases of uniformly
		disordered two-dimensional superconductors}.
	\newblock \emph{\bibinfo{journal}{Physical Review Letters}}
	\textbf{\bibinfo{volume}{69}}, \bibinfo{pages}{1604} (\bibinfo{year}{1992}).
	
	\bibitem{yazdani1995superconducting}
	\bibinfo{author}{Yazdani, A.} \& \bibinfo{author}{Kapitulnik, A.}
	\newblock \bibinfo{title}{Superconducting-insulating transition in
		two-dimensional a-moge thin films}.
	\newblock \emph{\bibinfo{journal}{Physical Review Letters}}
	\textbf{\bibinfo{volume}{74}}, \bibinfo{pages}{3037} (\bibinfo{year}{1995}).
	
	\bibitem{gantmakher2000observation}
	\bibinfo{author}{Gantmakher, V.}, \bibinfo{author}{Golubkov, M.},
	\bibinfo{author}{Dolgopolov, V.}, \bibinfo{author}{Shashkin, A.} \&
	\bibinfo{author}{Tsydynzhapov, G.}
	\newblock \bibinfo{title}{Observation of the parallel-magnetic-field-induced
		superconductor-insulator transition in thin amorphous ino films}.
	\newblock \emph{\bibinfo{journal}{JETP Letters}} \textbf{\bibinfo{volume}{71}},
	\bibinfo{pages}{473} (\bibinfo{year}{2000}).
	
	\bibitem{sambandamurthy2004superconductivity}
	\bibinfo{author}{Sambandamurthy, G.}, \bibinfo{author}{Engel, L.~W.},
	\bibinfo{author}{Johansson, A.} \& \bibinfo{author}{Shahar, D.}
	\newblock \bibinfo{title}{Superconductivity-related insulating behavior}.
	\newblock \emph{\bibinfo{journal}{Physical Review Letters}}
	\textbf{\bibinfo{volume}{92}}, \bibinfo{pages}{107005}
	(\bibinfo{year}{2004}).
	
	\bibitem{sambandamurthy2005experimental}
	\bibinfo{author}{Sambandamurthy, G.}, \bibinfo{author}{Engel, L.~W.},
	\bibinfo{author}{Johansson, A.}, \bibinfo{author}{Peled, E.} \&
	\bibinfo{author}{Shahar, D.}
	\newblock \bibinfo{title}{Experimental evidence for a collective insulating
		state in two-dimensional superconductors}.
	\newblock \emph{\bibinfo{journal}{Physical Review Letters}}
	\textbf{\bibinfo{volume}{94}}, \bibinfo{pages}{017003}
	(\bibinfo{year}{2005}).
	
	\bibitem{steiner2005possible}
	\bibinfo{author}{Steiner, M.~A.}, \bibinfo{author}{Boebinger, G.} \&
	\bibinfo{author}{Kapitulnik, A.}
	\newblock \bibinfo{title}{Possible field-tuned superconductor-insulator
		transition in high-t c superconductors: Implications for pairing at high
		magnetic fields}.
	\newblock \emph{\bibinfo{journal}{Physical Review Letters}}
	\textbf{\bibinfo{volume}{94}}, \bibinfo{pages}{107008}
	(\bibinfo{year}{2005}).
	
	\bibitem{baturina2005quantum}
	\bibinfo{author}{Baturina, T.~I.}, \bibinfo{author}{Bentner, J.},
	\bibinfo{author}{Strunk, C.}, \bibinfo{author}{Baklanov, M.~R.} \&
	\bibinfo{author}{Satta, A.}
	\newblock \bibinfo{title}{From quantum corrections to magnetic-field-tuned
		superconductor--insulator quantum phase transition in tin films}.
	\newblock \emph{\bibinfo{journal}{Physica B: Condensed Matter}}
	\textbf{\bibinfo{volume}{359}}, \bibinfo{pages}{500} (\bibinfo{year}{2005}).
	
	\bibitem{baturina2007quantum}
	\bibinfo{author}{Baturina, T.~I.}, \bibinfo{author}{Strunk, C.},
	\bibinfo{author}{Baklanov, M.~R.} \& \bibinfo{author}{Satta, A.}
	\newblock \bibinfo{title}{Quantum metallicity on the high-field side of the
		superconductor-insulator transition}.
	\newblock \emph{\bibinfo{journal}{Physical Review Letters}}
	\textbf{\bibinfo{volume}{98}}, \bibinfo{pages}{127003}
	(\bibinfo{year}{2007}).
	
	\bibitem{crane2007fluctuations}
	\bibinfo{author}{Crane, R.~W.}, \bibinfo{author}{Armitage, N.~P.}, \bibinfo{author}{Johansson, A.}, \bibinfo{author}{Sambandamurthy, G.}, \bibinfo{author}{Shahar, D.}, \& \bibinfo{author}{Gruner, G.}
	\newblock \bibinfo{title}{Fluctuations, dissipation, and nonuniversal
		superfluid jumps in two-dimensional superconductors}.
	\newblock \emph{\bibinfo{journal}{Physical Review B}}
	\textbf{\bibinfo{volume}{75}}, \bibinfo{pages}{094506}
	(\bibinfo{year}{2007}).
	
	\bibitem{vinokur2008superinsulator}
	\bibinfo{author}{Vinokur, V.~M.}, \bibinfo{author}{Baturina, T.~I.}, \bibinfo{author}{Fistul, M.~V.}, \bibinfo{author}{Mironov, A.~Yu.}, \bibinfo{author}{Baklanov, M.~R.}, \& \bibinfo{author}{Strunk, C.}
	\newblock \bibinfo{title}{Superinsulator and quantum synchronization}.
	\newblock \emph{\bibinfo{journal}{Nature}} \textbf{\bibinfo{volume}{452}},
	\bibinfo{pages}{613} (\bibinfo{year}{2008}).
	
	\bibitem{parendo2005electrostatic}
	\bibinfo{author}{Parendo, K.~A.}, \bibinfo{author}{Sarwa B. Tan, K.~H.}, \bibinfo{author}{Bhattacharya, A.}, \bibinfo{author}{Eblen-Zayas, M.}, \bibinfo{author}{Staley, N.~E.} \& \bibinfo{author}{Goldman, A.~M.}
	\newblock \bibinfo{title}{Electrostatic tuning of the superconductor-insulator
		transition in two dimensions}.
	\newblock \emph{\bibinfo{journal}{Physical Review Letters}}
	\textbf{\bibinfo{volume}{94}}, \bibinfo{pages}{197004}
	(\bibinfo{year}{2005}).
	
	\bibitem{mondal2011phase}
	\bibinfo{author}{Mondal, M.}, \bibinfo{author}{Kamlapure, .}, \bibinfo{author}{Chand,M.}, \bibinfo{author}{Saraswat, G.}, \bibinfo{author}{Kumar, S.}, \bibinfo{author}{Jesudasan, J.}, \bibinfo{author}{Benfatto, L.}, \bibinfo{author}{Tripathi, V.} \&
	\bibinfo{author}{Raychaudhuri, P.}
	\newblock \bibinfo{title}{Phase fluctuations in a strongly disordered s-wave
		nbn superconductor close to the metal-insulator transition}.
	\newblock \emph{\bibinfo{journal}{Physical Review Letters}}
	\textbf{\bibinfo{volume}{106}}, \bibinfo{pages}{047001}
	(\bibinfo{year}{2011}).
	
	\bibitem{higgs}
	\bibinfo{author}{Sherman, D.} \emph{et~al.}
	\newblock \bibinfo{title}{The higgs mode in disordered superconductors close to
		a quantum phase transition}.
	\newblock \emph{\bibinfo{journal}{Nature Physics}}  (\bibinfo{year}{2015}).
	
	\bibitem{poran2011disorder}
	\bibinfo{author}{Poran, S.}, \bibinfo{author}{Shimshoni, E.} \&
	\bibinfo{author}{Frydman, A.}
	\newblock \bibinfo{title}{Disorder-induced superconducting ratchet effect in
		nanowires}.
	\newblock \emph{\bibinfo{journal}{Physical Review B}}
	\textbf{\bibinfo{volume}{84}}, \bibinfo{pages}{014529}
	(\bibinfo{year}{2011}).
	
	\bibitem{shimshoni1998transport}
	\bibinfo{author}{Shimshoni, E.}, \bibinfo{author}{Auerbach, A.} \&
	\bibinfo{author}{Kapitulnik, A.}
	\newblock \bibinfo{title}{Transport through quantum melts}.
	\newblock \emph{\bibinfo{journal}{Physical Review Letters}}
	\textbf{\bibinfo{volume}{80}}, \bibinfo{pages}{3352} (\bibinfo{year}{1998}).
	
	\bibitem{bouadim2011single}
	\bibinfo{author}{Bouadim, K.}, \bibinfo{author}{Loh, Y.~L.},
	\bibinfo{author}{Randeria, M.} \& \bibinfo{author}{Trivedi, N.}
	\newblock \bibinfo{title}{Single-and two-particle energy gaps across the
		disorder-driven superconductor-insulator transition}.
	\newblock \emph{\bibinfo{journal}{Nature Physics}}
	\textbf{\bibinfo{volume}{7}}, \bibinfo{pages}{884} (\bibinfo{year}{2011}).
	
	\bibitem{dubi2007nature}
	\bibinfo{author}{Dubi, Y.}, \bibinfo{author}{Meir, Y.} \&
	\bibinfo{author}{Avishai, Y.}
	\newblock \bibinfo{title}{Nature of the superconductor--insulator transition in
		disordered superconductors}.
	\newblock \emph{\bibinfo{journal}{Nature}} \textbf{\bibinfo{volume}{449}},
	\bibinfo{pages}{876} (\bibinfo{year}{2007}).
	
	\bibitem{erez2010thermal}
	\bibinfo{author}{Erez, A.} \& \bibinfo{author}{Meir, Y.}
	\newblock \bibinfo{title}{Thermal phase transition in two-dimensional
		disordered superconductors}.
	\newblock \emph{\bibinfo{journal}{EPL}} \textbf{\bibinfo{volume}{91}},
	\bibinfo{pages}{47003} (\bibinfo{year}{2010}).
	
	\bibitem{erez2013proposed}
	\bibinfo{author}{Erez, A.} \& \bibinfo{author}{Meir, Y.}
	\newblock \bibinfo{title}{Proposed measurement of spatial correlations at the
		berezinski-kosterlitz-thouless transition of superconducting thin films}.
	\newblock \emph{\bibinfo{journal}{Physical Review Letters}}
	\textbf{\bibinfo{volume}{111}}, \bibinfo{pages}{187002}
	(\bibinfo{year}{2013}).
	
	\bibitem{cohen1969effect}
	\bibinfo{author}{Cohen, R.~W.}, \bibinfo{author}{Abeles, B.} \&
	\bibinfo{author}{Fuselier, C.}
	\newblock \bibinfo{title}{Effect of fluctuations in the superconducting order
		parameter on the tunneling density of states}.
	\newblock \emph{\bibinfo{journal}{Physical Review Letters}}
	\textbf{\bibinfo{volume}{23}}, \bibinfo{pages}{377} (\bibinfo{year}{1969}).
	
	\bibitem{abrahams1970effect}
	\bibinfo{author}{Abrahams, E.}, \bibinfo{author}{Redi, M.} \&
	\bibinfo{author}{Woo, J.~W.}
	\newblock \bibinfo{title}{Effect of fluctuations on electronic properties above
		the superconducting transition}.
	\newblock \emph{\bibinfo{journal}{Physical Review B}}
	\textbf{\bibinfo{volume}{1}}, \bibinfo{pages}{208} (\bibinfo{year}{1970}).
	
	\bibitem{di1990superconductive}
	\bibinfo{author}{Di~Castro, C.}, \bibinfo{author}{Raimondi, R.},
	\bibinfo{author}{Castellani, C.} \& \bibinfo{author}{Varlamov, A.~A.}
	\newblock \bibinfo{title}{Superconductive fluctuations in the density of states
		and tunneling resistance in high-$T_c$ superconductors}.
	\newblock \emph{\bibinfo{journal}{Physical Review B}}
	\textbf{\bibinfo{volume}{42}}, \bibinfo{pages}{10211} (\bibinfo{year}{1990}).
	
	\bibitem{varlamov1999role}
	\bibinfo{author}{Varlamov, A.}, \bibinfo{author}{Balestrino, G.},
	\bibinfo{author}{Milani, E.} \& \bibinfo{author}{Livanov, D.}
	\newblock \bibinfo{title}{The role of density of states fluctuations in the
		normal state properties of high t c superconductors}.
	\newblock \emph{\bibinfo{journal}{Advances in Physics}}
	\textbf{\bibinfo{volume}{48}}, \bibinfo{pages}{655--783}
	(\bibinfo{year}{1999}).
	
	\bibitem{Dynes}
	\bibinfo{author}{Dynes, R.~C.}, \bibinfo{author}{Narayanamurti, V.} \&
	\bibinfo{author}{Garno, J.~P.}
	\newblock \bibinfo{title}{Direct measurement of quasiparticle-lifetime
		broadening in a strong-coupled superconductor}.
	\newblock \emph{\bibinfo{journal}{Phys. Rev. Lett.}}
	\textbf{\bibinfo{volume}{41}}, \bibinfo{pages}{1509} (\bibinfo{year}{1978}).
	
	\bibitem{BCS}
	\bibinfo{author}{Bardeen, J.}, \bibinfo{author}{Cooper, L.~N.} \&
	\bibinfo{author}{Schrieffer, J.~R.}
	\newblock \bibinfo{title}{Theory of superconductivity}.
	\newblock \emph{\bibinfo{journal}{Phys. Rev.}} \textbf{\bibinfo{volume}{108}},
	\bibinfo{pages}{1175} (\bibinfo{year}{1957}).
	
	\bibitem{barber1994tunneling}
	\bibinfo{author}{Barber~Jr, R.~P.}, \bibinfo{author}{Merchant, L.~M.},
	\bibinfo{author}{La~Porta, A.} \& \bibinfo{author}{Dynes, R.~C.}
	\newblock \bibinfo{title}{Tunneling into granular pb films in the
		superconducting and insulating regimes}.
	\newblock \emph{\bibinfo{journal}{Physical Review B}}
	\textbf{\bibinfo{volume}{49}}, \bibinfo{pages}{3409} (\bibinfo{year}{1994}).
	
	\bibitem{sherman2012measurement}
	\bibinfo{author}{Sherman, D.}, \bibinfo{author}{Kopnov, G.},
	\bibinfo{author}{Shahar, D.} \& \bibinfo{author}{Frydman, A.}
	\newblock \bibinfo{title}{Measurement of a superconducting energy gap in a
		homogeneously amorphous insulator}.
	\newblock \emph{\bibinfo{journal}{Physical Review Letters}}
	\textbf{\bibinfo{volume}{108}}, \bibinfo{pages}{177006}
	(\bibinfo{year}{2012}).
	
	\bibitem{sacepe2011localization}
	\bibinfo{author}{Sac{\'e}p{\'e}, B.}, \bibinfo{author}{Dubouchet, T.}, \bibinfo{author}{Chapelier, C.}, \bibinfo{author}{Sanquer, M.}, \bibinfo{author}{Ovadia, M.}, \bibinfo{author}{Shahar, D.}, \bibinfo{author}{Feigel'man, M.} \& \bibinfo{author}{Ioffe, L.}
	\newblock \bibinfo{title}{Localization of preformed cooper pairs in disordered
		superconductors}.
	\newblock \emph{\bibinfo{journal}{Nature Physics}}
	\textbf{\bibinfo{volume}{7}}, \bibinfo{pages}{239} (\bibinfo{year}{2011}).
	
	\bibitem{zasadzinski1992tunneling}
	\bibinfo{author}{Zasadzinski, J.}, \bibinfo{author}{Tralshawala, N.}, \bibinfo{author}{Romano, P.}, \bibinfo{author}{Huang, Q.}, \bibinfo{author}{Chen, Jun.} \& \bibinfo{author}{Gray, K.~E.}
	\newblock \bibinfo{title}{Tunneling density of states in cuprate and bismuthate
		superconductors}.
	\newblock \emph{\bibinfo{journal}{Journal of Physics and Chemistry of Solids}}
	\textbf{\bibinfo{volume}{53}}, \bibinfo{pages}{1635} (\bibinfo{year}{1992}).
	
	\bibitem{dynes1994order}
	\bibinfo{author}{Dynes, R.}
	\newblock \bibinfo{title}{The order parameter of high tc superconductors;
		experimental probes}.
	\newblock \emph{\bibinfo{journal}{Solid state communications}}
	\textbf{\bibinfo{volume}{92}}, \bibinfo{pages}{53} (\bibinfo{year}{1994}).
	
	\bibitem{norman1998phenomenology}
	\bibinfo{author}{Norman, M.~R.}, \bibinfo{author}{Randeria, M.},
	\bibinfo{author}{Ding, H.} \& \bibinfo{author}{Campuzano, J.~C.}
	\newblock \bibinfo{title}{Phenomenology of the low-energy spectral function in
		high-$T_c$ superconductors}.
	\newblock \emph{\bibinfo{journal}{Physical Review B}}
	\textbf{\bibinfo{volume}{57}}, \bibinfo{pages}{R11093}
	(\bibinfo{year}{1998}).
	
	\bibitem{norman2001temperature}
	\bibinfo{author}{Norman, M.}, \bibinfo{author}{Kaminski, A.},
	\bibinfo{author}{Mesot, J.} \& \bibinfo{author}{Campuzano, J.}
	\newblock \bibinfo{title}{Temperature evolution of the spectral peak in
		high-temperature superconductors}.
	\newblock \emph{\bibinfo{journal}{Physical Review B}}
	\textbf{\bibinfo{volume}{63}}, \bibinfo{pages}{140508}
	(\bibinfo{year}{2001}).
	
	\bibitem{norman2007modeling}
	\bibinfo{author}{Norman, M.~R.}, \bibinfo{author}{Kanigel, A.},
	\bibinfo{author}{Randeria, M.}, \bibinfo{author}{Chatterjee, U.} \&
	\bibinfo{author}{Campuzano, J.~C.}
	\newblock \bibinfo{title}{Modeling the fermi arc in underdoped cuprates}.
	\newblock \emph{\bibinfo{journal}{Physical Review B}}
	\textbf{\bibinfo{volume}{76}}, \bibinfo{pages}{174501}
	(\bibinfo{year}{2007}).
	
	\bibitem{mahan2000many}
	\bibinfo{author}{Mahan, G.}
	\newblock \emph{\bibinfo{title}{Many-Particle Physics}}.
	\newblock Physics of Solids and Liquids (\bibinfo{publisher}{Springer},
	\bibinfo{year}{2000}).
	
	\bibitem{altland2006condensed}
	\bibinfo{author}{Altland, A.} \& \bibinfo{author}{Simons, B.}
	\newblock \emph{\bibinfo{title}{Condensed Matter Field Theory}}
	(\bibinfo{publisher}{Cambridge University Press}, \bibinfo{year}{2006}).
	
	\bibitem{burmistrov2016local}
	\bibinfo{author}{Burmistrov, I.~S.}, \bibinfo{author}{Gornyi, I.~V.} \&
	\bibinfo{author}{Mirlin, A.~D.}
	\newblock \bibinfo{title}{Local density of states and its mesoscopic
		fluctuations near the transition to a superconducting state in disordered
		systems}.
	\newblock \emph{\bibinfo{journal}{Physical Review B}}
	\textbf{\bibinfo{volume}{93}}, \bibinfo{pages}{205432}
	(\bibinfo{year}{2016}).
	
	\bibitem{Anderson}
	\bibinfo{author}{Anderson, P.~W.}
	\newblock \bibinfo{title}{Theory of dirty superconductors}.
	\newblock \emph{\bibinfo{journal}{Journal of Physics and Chemistry of Solids}}
	\textbf{\bibinfo{volume}{11}}, \bibinfo{pages}{26} (\bibinfo{year}{1959}).
	
	\bibitem{seibold2015amplitude}
	\bibinfo{author}{Seibold, G.}, \bibinfo{author}{Benfatto, L.},
	\bibinfo{author}{Castellani, C.} \& \bibinfo{author}{Lorenzana, J.}
	\newblock \bibinfo{title}{Amplitude, density, and current correlations of
		strongly disordered superconductors}.
	\newblock \emph{\bibinfo{journal}{Physical Review B}}
	\textbf{\bibinfo{volume}{92}}, \bibinfo{pages}{064512}
	(\bibinfo{year}{2015}).
	
	\bibitem{sondhi1997continuous}
	\bibinfo{author}{Sondhi, S.}, \bibinfo{author}{Girvin, S.},
	\bibinfo{author}{Carini, J.} \& \bibinfo{author}{Shahar, D.}
	\newblock \bibinfo{title}{Continuous quantum phase transitions}.
	\newblock \emph{\bibinfo{journal}{Reviews of Modern Physics}}
	\textbf{\bibinfo{volume}{69}}, \bibinfo{pages}{315} (\bibinfo{year}{1997}).
	
	
	\bibitem{boyer2007imaging}
	\bibinfo{author}{Boyer, M.~C.}, \bibinfo{author}{Wise, W.~D.}, \bibinfo{author}{Chatterjee, K.}, \bibinfo{author}{Yi, M.}, \bibinfo{author}{Kondo, T.}, \bibinfo{author}{Takeuchi, T.}, \bibinfo{author}{Ikuta, H.}, \& \bibinfo{author}{Hudson, E.~W.},
	\newblock \bibinfo{title}{Imaging the two gaps of the high-temperature
		superconductor ${\mathrm{Bi_2Sr_2CuO_{6+x}}}$}.
	\newblock \emph{\bibinfo{journal}{Nature Physics}}
	\textbf{\bibinfo{volume}{3}}, \bibinfo{pages}{802} (\bibinfo{year}{2007}).
	
	\bibitem{fujita2011spectroscopic}
	\bibinfo{author}{Fujita, K.}, \bibinfo{author}{Schmidt, A.~R.}, \bibinfo{author}{Kim, E.}, \bibinfo{author}{Lawler, M.~J.}, \bibinfo{author}{Lee, D.~H.}, \bibinfo{author}{Davis, J.~C.}, \bibinfo{author}{Hiroshi, E.}, \& \bibinfo{author}{Uchida, S.},
	\newblock \bibinfo{title}{Spectroscopic imaging scanning tunneling microscopy
		studies of electronic structure in the superconducting and pseudogap phases
		of cuprate high-$T_c$ superconductors}.
	\newblock \emph{\bibinfo{journal}{Journal of the Physical Society of Japan}}
	\textbf{\bibinfo{volume}{81}}, \bibinfo{pages}{011005}
	(\bibinfo{year}{2011}).
	
\end{thebibliography}

\newpage

\newpage

\section{SUPPLEMENTAL MATERIALS}

\renewcommand{\theequation}{S\arabic{equation}}   
\setcounter{equation}{0}

\renewcommand{\thefigure}{S\arabic{figure}}   
\setcounter{figure}{0}

\renewcommand{\thesection}{SI-\arabic{section}}   
\setcounter{section}{0}

\renewcommand{\thetable}{S\arabic{table}}   
\setcounter{table}{0}

\section{Time-dependent superconducting fluctuations}	

The general form of the pairing fluctuation function, as presented in Eq.~(\ref{pairing1}), contains an integral over the frequency $\Omega$. In the main text we considered correlation functions of the form $C(\textbf{q}, \Omega) = C(\textbf{q})\delta(\Omega)$, for which the integral has a finite contribution only for $\Omega = 0$. In this appendix we show numerically that our main findings are not changed significantly if the frequency dependence of $C$ is explicitly taken into account.

\begin{figure}[h]
	\includegraphics[scale=0.75]{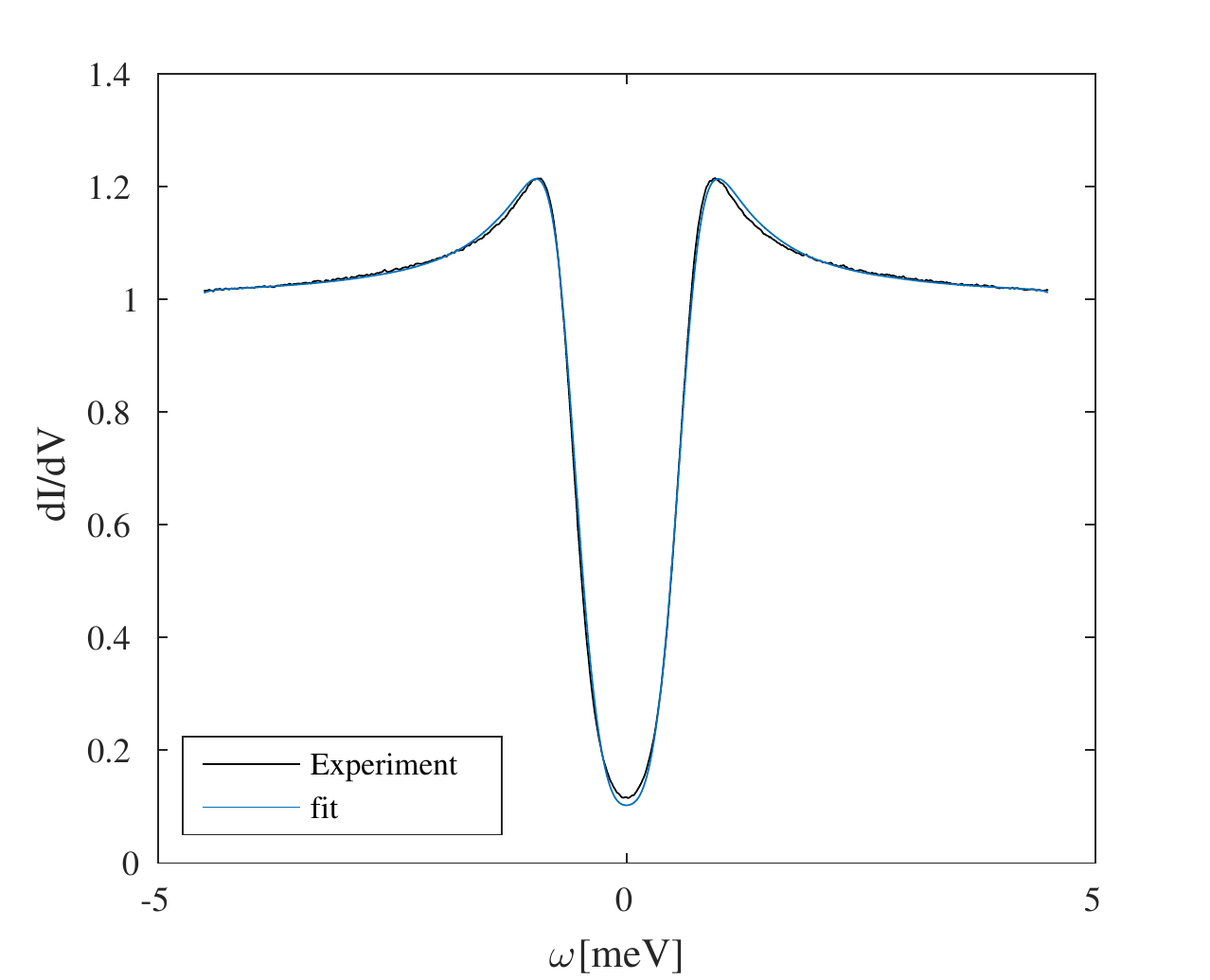}
	\centering
	\caption{Optimal fit of the DOS (Eq.~\ref{DOS}) using the sum of short range and time-dependent quasi long range superconducting fluctuations (Eqs.~(\ref{eq:DDSR}) and (\ref{COmega}) respectively).}
	\label{FitOmega1}
\end{figure}

Following the assumptions mentioned in the main text, we approximate $\varepsilon_{\textbf{k}-\textbf{q}} \approx -\textbf{k}_{F}\cdot \textbf{q} = -v_{F}q\cos(\theta)$, where $\theta$ is the angle between $\textbf{k}_{F}$ and $\textbf{q}$, and obtain
\begin{align}\label{ThetaInt}
\mathcal{D}^2(\omega)=&\int_{0}^{2\pi} \int_{0}^{\infty}\int_{-\infty}^{\infty} d\Omega ~dq ~d\theta \times \nonumber\\&\dfrac{\omega q}{i\Omega+\omega -v_{F}q\cos(\theta)} C(q,\Omega).
\end{align}
The integral in Eq.~(\ref{ThetaInt}) can be evaluated using the variable substitution  $t=\tan(\frac{\theta}{2})$, for which  $\cos(\theta) = \frac{1-t^2}{1+t^2}$ and $d\theta = \frac{2}{1+t^2}dt$. Also, one needs to assume that $C(\textbf{q}, \Omega)$ is $\theta$ independent, i.e. depends only on the absolute value of $\textbf{q}$. 
By introducing $A= q \omega C(q,\Omega), a=i\Omega+\omega$, and $b=v_{F}q$, we can rewrite the integral over $\theta$ as
\begin{multline}
\int_{-\infty}^{\infty} dt \dfrac{2A}{t^2(a+b)+a-b} 
= \\
\int_{-\infty}^{\infty} dt \dfrac{2A}{a+b}\dfrac{1}{(t+i\sqrt{m})(t-i\sqrt{m})},
\end{multline}
where $m = \frac{a-b}{a+b}$.
By performing a standard Cauchy contour integral, we obtain 
\begin{align}
\dfrac{2\pi A}{\sqrt{(a+b)(a-b)}} \sign \left({\rm Re} \sqrt{\dfrac{a-b}{a+b}}\right) .
\end{align}
Recalling the original definition of $A$, $a$ and $b$, Eq.~(\ref{pairing1}) can be written now as
\begin{multline}\label{DOmega}
\mathcal{D}^2(\omega) =  \int_{-\infty}^{\infty} \int_{0}^{\infty} dq ~d\Omega \times \\ \dfrac{2\pi q \omega C(q,\Omega)}{\sqrt{(i\Omega +\omega+v_{F}q)(i\Omega +\omega-v_{F}q)}} \times \\
\sign\left( {\rm Re} \sqrt{\dfrac{i\Omega +\omega-v_{F}q}{i\Omega +\omega+v_{F}q}} \right).
\end{multline}
To study the effects of a finite velocity of the collective excitations $v\neq0$, we consider QLR fluctuations described by

\begin{align}\label{COmega}
C_{QLR}(q, \Omega) = \frac{A}{v^2q^2+v^2q_{QLR}^2+\Omega^2}.
\end{align}
Here $v$ is the collective mode velocity and $A$ a normalization parameter . We next numerically compute $\DD(\omega)$ from Eq. (\ref{DOmega}), add the contributions from the SR correlations Eq. (11), and compare the resulting differential conductivity with the experiments. Using a derivative free optimization algorithm, we then minimize the $\chi^2$ difference between theory and experiment with respect to $\gamma, A, v, q_0$. The result of this minimization procedure gives an excellent agreement between theory and experiment $(\chi^2=0.007)$: as shown in Fig.~\ref{FitOmega1} the two curves are almost indistinguishable. This optimal fit is obtained for $k_{F}v = 3.7$meV$ \ll k_{F}v_F \sim 1$eV, confirming that $v \ll v_F$, as expected. This observation justifies the approximation  $C(\textbf{q},\Omega) \approx C(\textbf{q})\delta(\Omega)$ used in the main text. 

\newpage
\begin{widetext}
\section{Derivation of the DOS}
In the main text we introduced the Hamiltonian $H=H_{0}+H_{\Delta}$, where $H_{0} = \sum_{\textbf{k}, \sigma} \varepsilon_{\textbf{k}} c^{\dagger}_{\textbf{k}, \sigma}c_{\textbf{k}, \sigma}$ describes free electrons (quasiparticles) with a Fermi surface at $\varepsilon_{\textbf{k}}=0$. We represented superconducting fluctuations by a randomly fluctuating bosonic field $\Delta(\textbf{r}, t)$ coupled to the fermions by
\begin{align}
H_{\Delta} = \Delta(\textbf{r},t)c_{\uparrow}^\dagger (\textbf{r},t) c_{\downarrow}^\dagger(\textbf{r},t) + {\rm H. c.}\;.
\end{align}

In Nambu space, the bare Green's function corresponding to the free Hamiltonian is 
\begin{align}
G^{-1}_{0}(\textbf{k},\omega) = \begin{pmatrix} i\omega-\varepsilon_{\textbf{k}} \ \ \ \  \ \ 0 \\ 0 \ \ \ \ \ i\omega+\varepsilon_{\textbf{k}} \end{pmatrix},
\end{align}	
where $\omega \equiv \omega_{n}$, is the fermionic Matsubara frequency. The interaction matrix in Nambu space, which we denote by $V(\textbf{k}-\textbf{k}',\omega-\omega')$, is define according to $H_{\Delta}$ as:

\begin{align}\label{Vdef}
V(\textbf{k}-\textbf{k}',\omega- \omega') =\begin{pmatrix}
0 \ \ \  \ \ \ \Delta(\textbf{k}-\textbf{k}', \omega-\omega') \\ \Delta^{*}(\textbf{k}-\textbf{k}', \omega-\omega')  \ \  \ \  \ \   0 \end{pmatrix}.
\end{align}

Using the imaginary time path integrals formulation, and performing a standard resummation to obtain
\begin{align}\label{GR}
\mathcal{G}(\textbf{k},\omega) =  \left[G^{-1}_{0}-\Sigma\right]^{-1},
\end{align}
where $\Sigma = \sum_{\textbf{k}',\textbf{k}'',\omega',\omega''} V(\textbf{k}-\textbf{k}',\omega-\omega')G_{0}(\textbf{k}',\omega')V(\textbf{k}'-\textbf{k}'',\omega'-\omega'')$ is the bubble of Fig.~\ref{Fyenman}. We multiply both the numerator and denominator of Eq.~(\ref{GR}) by $G_{0}(\textbf{k},\omega)$ to obtain	
\begin{align}\label{GR2}
\mathcal{G}(\textbf{k},\omega)=  \dfrac{G_{0}(\textbf{k},\omega)}{I-\sum_{\textbf{k}',\textbf{k}'',\omega',\omega''}  V(\textbf{k}-\textbf{k}',\omega-\omega')G_{0}(\textbf{k}',\omega')V(\textbf{k}''-\textbf{k}',\omega''-\omega')G_{0}(\textbf{k},\omega)},
\end{align}
where $I$ is the 2X2 unit matrix. Using the defenition of the bare Green's function and interaction matrix, Eqs.~(\ref{Vdef},\ref{GR}), the multiplication of $V(\textbf{k}-\textbf{k}',\omega-\omega') G_{0}(\textbf{k}',\omega')$ gives

\begin{align}\label{VG}
\begin{pmatrix}
0  & \dfrac{\Delta(\textbf{k}-\textbf{k}', \omega-\omega')}{(i\omega'+\varepsilon_{\textbf{k}'})} \\ \dfrac{\Delta^{*}(\textbf{k}-\textbf{k}', \omega-\omega')}{(i\omega'-\varepsilon_{\textbf{k}'})}  & 0 \end{pmatrix}.
\end{align}
The  second multiplication is performed equivalently, where $\Delta(\textbf{k}-\textbf{k}',\omega-\omega') \to \Delta(\textbf{k}''-\textbf{k}',\omega''-\omega')$ and $(i\omega' \pm \varepsilon_{\textbf{k}}') \to (i\omega \pm \varepsilon_{\textbf{k}})$.

Substituting Eq.~(\ref{VG}) into Eq.~(\ref{GR2}) and taking the expectation value, we obtain:
\begin{align}
\mathcal{G}(\textbf{k},\omega) = G_{0}(\textbf{k},\omega)\left[I-\sum_{\textbf{k}',\omega'}  \begin{pmatrix}
\dfrac{C(\textbf{k}-\textbf{k}', \omega-\omega')}{(i\omega-\varepsilon_{\textbf{k}})(i\omega'+\varepsilon_{\textbf{k}'})} & 0 \\ 0 & \dfrac{C(\textbf{k}-\textbf{k}', \omega-\omega')}{(i\omega+\varepsilon_{\textbf{k}})(i\omega'-\varepsilon_{\textbf{k}'})} \end{pmatrix}\right]^{-1},
\end{align}
where $C(\textbf{k}-\textbf{k}', \omega-\omega')\delta(\textbf{k}-\textbf{k}'')\delta(\omega-\omega'') = \langle\Delta(\textbf{k}-\textbf{k}',\omega'-\omega'') \Delta^{\star}(\textbf{k}''-\textbf{k}',\omega'-\omega'') \rangle $ is the correlation function of the field $\Delta$. We can further simplify this expression as 

\begin{align}
\mathcal{G}(\textbf{k},\omega)=G_{0}(\textbf{k},\omega)\begin{pmatrix}
1-\sum_{\textbf{q},\Omega}\dfrac{C(\textbf{q},\Omega)}{[i\omega-\varepsilon_{\textbf{k}}][i(\omega-\Omega)+\varepsilon_{\textbf{k}-\textbf{q}}]} & 0 \\ 0 & 1- \sum_{\textbf{q},\Omega}\dfrac{C(\textbf{q},\Omega)}{[i\omega+\varepsilon_{\textbf{k}}][i(\omega-\Omega)-\varepsilon_{\textbf{k}-\textbf{q}}]}
\end{pmatrix}^{-1},
\end{align}
where we define $\textbf{k}-\textbf{k}'=\textbf{q}$ and $\omega-\omega'=\Omega$.
We now invert the matrix and multiply it by $G_{0}(\textbf{k},\omega)$. Then, we perform an analytic continuation $i\omega \to \omega+i0^{+}$ and write the final result in the form of the retarded Green's function
\begin{align}
G^{ret}(\textbf{k},\omega) = \begin{pmatrix}
\dfrac{1}{\omega+i0^{+}-\varepsilon_{\textbf{k}}-\sum_{\textbf{q},\omega}\dfrac{C(\textbf{q},\Omega)}{\omega+i0^{+}-i\Omega+\varepsilon_{\textbf{k}-\textbf{q}}}} & 0 \\ \\ 0 & \dfrac{1}{\omega+i0^{+}+\varepsilon_{\textbf{k}}-\sum_{\textbf{q},\omega}\dfrac{C(\textbf{q},\Omega)}{\omega+i0^{+}-i\Omega-\varepsilon_{\textbf{k}-\textbf{q}}}}
\end{pmatrix}.
\end{align}
In order to write $G^{ret}(\textbf{k},\omega)$ in a way which is similar to Dynes formula, we multiply the numerator and denominator of $G^{ret}_{1,1}$ and $G^{ret}_{2,2}$ by $(\omega+i0^{+})\pm\varepsilon_{\textbf{k}}$ respectively. As a result \begin{align}\label{G11}
G^{ret}_{1,1}(\textbf{k},\omega) =  \dfrac{\omega+i0^{+}+\varepsilon_{\textbf{k}}}{(\omega+i0^{+})^2-\varepsilon^{2}_{\textbf{k}}-\sum_{\textbf{q},\Omega}\dfrac{(\omega+i0^{+}+\varepsilon_{\textbf{k}})C(\textbf{q},\Omega)}{\omega+i0^{+}-i\Omega+\varepsilon_{\textbf{k}-\textbf{q}}}},
\end{align}
and 
\begin{align}\label{G22}
G^{ret}_{2,2}(\textbf{k},\omega) = \dfrac{\omega+i0^{+}-\varepsilon_{\textbf{k}}}{(\omega+i0^{+})^2-\varepsilon^{2}_{\textbf{k}}- \sum_{\textbf{q},\Omega}\dfrac{(\omega-\varepsilon_{\textbf{k}})C(\textbf{q},\Omega)}{\omega+i0^{+}-i\Omega)-\varepsilon_{\textbf{k}-\textbf{q}}}}.
\end{align}

The last term on each denominator was define in the main text as $\mathcal{D}^2$, and evaluated at $\textbf{k}_{F}$ where $\varepsilon_{\textbf{k}_{F}} = 0$ (the justification can be found there). Thus, the two expressions are the same, except the sign of $\varepsilon_{\textbf{k}-\textbf{q}}$ in the denominator of $\mathcal{D}^2$. We approximate $\varepsilon_{\textbf{k}_{F}-\textbf{q}}\approx -(v_F/k_F) (\textbf{k}_{F} \cdot \textbf{q})$ and take $\textbf{q}\to -\textbf{q}$ in Eq.~(\ref{G22}), so the denominator of both expression is equal. Consequently, after one perform an analytic continuation $i\omega \to \omega+i0^{+}$ the trace can be written as
\begin{align}
\rm Tr[G^{ret}(\textbf{k},\omega)]= \dfrac{\omega+i0^{+}}{(\omega+i0^{+})^2-\varepsilon_{\textbf{k}}^2-\mathcal{D}^2(\omega)}
\end{align}
where 
\begin{align}\label{D}
\mathcal{D}^2(\omega) = \int d^{2}\textbf{q}\int d\Omega \dfrac{\omega+i0^{+}}{i\Omega+\omega+i0^{+}+\varepsilon_{\textbf{k}_{F}-\textbf{q}}} C(\textbf{q},\Omega).
\end{align}

\end{widetext}
\end{document}